\documentclass[a4paper]{llncs}
\usepackage{amssymb}
\usepackage{graphicx}
\usepackage[latin1]{inputenc}
\usepackage{rotating}						
\usepackage{multirow}
\usepackage{booktabs}           

\usepackage{url}
\urldef{\mailsa}\path|pedrohgabriel@gmail.com,|
\urldef{\mailsb}\path|{miguel.goulao,vasco.amaral}@di.fct.unl.pt|
\newcommand{\keywords}[1]{\par\addvspace\baselineskip
\noindent\keywordname\enspace\ignorespaces#1}

\begin{document}

\mainmatter  

\title{Do Software Languages Engineers Evaluate their Languages?}

\titlerunning{}
%
%
\author{Pedro Gabriel \and Miguel Goulão\and Vasco Amaral}
%
\authorrunning{}

\institute{CITI, Departamento de Informática, Faculdade de Ciências e Tecnologia, FCT,\\ 
Universidade Nova de Lisboa, 2829-516 Caparica, Portugal
\\
\mailsa
\mailsb\\
\url{http://citi.di.fct.unl.pt/}}

%
%

\maketitle

\begin{abstract}
Domain Specific Languages (DSLs) can contribute to increment productivity, while reducing the required maintenance and programming expertise. We hypothesize that Software Languages Engineering (SLE) developers consistently skip, or relax, Language Evaluation. Based on the experience of engineering other types of software products, we assume that this may potentially lead to the deployment of inadequate languages. The fact that the languages already deal with concepts from the problem domain, and not the solution domain, is not enough to validate several issues at stake, such as its expressiveness, usability, effectiveness, maintainability, or even the domain expert's productivity while using them. We present a systematic review on articles published in top ranked venues, from 2001 to 2008, which report DSLs' construction, to characterize the common practice. This work confirms our initial hypothesis and lays the ground for the discussion on how to include a systematic approach to DSL evaluation in the SLE process.
\keywords{Domain Specific Languages, Systematic Review, Usability Evaluation, Experimental Software Engineering, Language Engineering Process Model}
\end{abstract}

\section{Introduction}
\label{sec:Introduction}
\vspace*{-.25cm}
Domain-driven development is an approach to software development which relies on Domain Specific Languages (DSLs) and Models (DSMs) to raise the level of abstraction, while at the same time narrowing down the design space \cite{Gray2004JVLC}. Among other claims, this shift of developers' focus to using abstractions that are part of the domain world, rather than general purpose abstractions closer to the design and code world, is said to bring important productivity gains, increased time-to-market responsiveness, and smaller training time, when compared to software development using general purpose design and programming languages \cite{Kelly2000ME}. The rationale is that developers no longer need to make error-prone mappings from domain concepts to design concepts, and onto programming language concepts. Instead, they can work directly with domain concepts. As such, domain experts can understand, validate, and modify the produced software, by adapting the domain-specific specifications \cite{Deursen1998JSM}. This approach relies on the existence of appropriate DSLs, which have to be built for each particular domain. Building such languages is a key challenge for software language engineers.

Software Languages Engineering (SLE) is becoming a mature and systematic activity, building upon the collective experience of a growing community, and the increasing availability of supporting tools. A typical SLE process starts with the Domain Engineering phase, in order to elicit the domain concepts. The following step is to design the language, capturing the referred concepts and their relationships. Then, the language is implemented, typically using workbench tools such as MetaEdit \cite{Smolander1991CAiSE}, MetaEdit+ \cite{Kelly1996CAiSE}, GMF/EMF \cite{Moore2004Book}, GME \cite{Vanderbilt2007} or Microsoft DSL Tools \cite{Cook2007Book}, and documented. A typical development process goes on to the testing, deployment, evolution, recovery, and retirement of languages. However streamlined the process is becoming, it still presents a serious gap in what should be a crucial phase: the Language Evaluation, or testing phase.

As with any other software product, we need to assure that the DSL is adequate to the end user (the Domain Expert). This covers not only the language's correctness, but also quality attributes, such as the language's usability, the maintainability of the produced systems, or the productivity of the developers using the DSL. A good DSL is hard to build because, as noted by Mernik et al. \cite{Mernik2005CSUR}, it requires both domain knowledge and language development expertise, and few people have both. We should assert claims like that the newly designed language brings efficiency to the process, or that it is usable and effective, with an unbiased evaluation process. Furthermore, we should be able to quantify the extent to which DSL's introduction brings economic benefits to organizations. Intuitively, we expect to observe a positive impact on the maintainability of systems specified with DSLs, when compared to those specified with general-purpose languages. We can find anecdotal reports of 3-10 times productivity improvements (e.g. \cite{Kelly2000ME,Weiss1999SPL,MetaCase2007Nokia}) or \textit{``clearly boosted development speeds''} \cite{MetaCase2007EADS} in industrial contexts. Unfortunately, it is not possible to make a fair meta-analysis on these claims. They are too vague and supported by testimonials by project managers rather than by detailed data that can be independently verified and used. What is a typical improvement, in quantifiable terms, particularly considering that we must take into account the development and maintainability of the DSL, and its set of supporting tools, as well? What is the impact of using several DSLs in an organization, each with its own maintainability cost attached? What is the break-even point from which it pays off to use a DSL? These questions are relevant, particularly because we find references to studies comparing alternative approaches to DSL construction with a major concern on the costs involved \cite{Kosar2008JIST,White2009IEEESoftware}. In short, when should we choose to develop and then adopt a DSL? We are able to find qualitative answers to this question \cite{Mernik2005CSUR}, but not quantitative-based ones (although Weiss and Lai do propose a formula for computing that break-even point \cite{Weiss1999SPL}). 

There are a few exceptions to this state of practice. Batory \textit{et al.} report on extensibility and maintainability improvements \cite{Batory2002TOSEM} brought by a combination of DSLs and Software Product Lines (SPLs), although it is unclear which share of the merit should be attributed to DSLs and which should be credited to SPLs. This is a potential confounding factor which can also be identified in most of the improvement claims concerning DSLs, as they are often used in combination with SPLs. Kieburtz \textit{et al.} report a series of experiments comparing favorably the usage of DSLs \textit{vs.} templates in code generation, with respect to flexibility, productivity, reliability and usability \cite{Kieburtz1996ICSE}. Hermans \textit{et al.} conducted a case study where the reliability and development costs were improved with the introduction of DSLs \cite{Hermans2009Models}. 
In the context of general purpose languages, we can find experimental comparative analysis of the productivity achieved by practitioners using different languages (e.g. \cite{Prechelt2000IEEEComputer}).

The software industry, in general, does not seem to invest much on the evaluation of DSLs. Among other possible explanations, this state of practice may stem from a lack of enough software experts that completely understand the SLE process, or from a lack of experimental evidence that clearly backs up the qualitative improvement claims that we often find in the literature. Without such evidence, it may be the case that decision makers consider proper language evaluation as a waste of time and resources. If so, they may prefer to risk using or selling inadequate DSLs rather than evaluating them properly. The incremental nature of a typical DSL life cycle may also give the erroneous feeling that the language is being implicitly validated due to the intense interaction with the domain experts. The problem there is that the domain experts involved in the language development may not be the end users, and may therefore introduce biases in the perception of the language design and usability. 

We present a systematic literature review to assess whether or not we can find evidence in the literature to back up our hypothesis: in general, software language engineers do not evaluate their languages with respect to their impact in the software development process in which the DSLs will be integrated. Systematic reviews use \textit{``a well-defined methodology to identify, analyse and interpret all available evidence related to a specific research question in a way that is unbiased and (to a degree) repeatable''} \cite{Kitchenham2007SR}.
To the best of our knowledge, there is no available systematic review and meta-analysis on the level of evaluation of DSLs reported in the literature. The review presented in this paper aims to fill in this gap. Ultimately, we aim to raise the community's awareness to the problem of poor validation of DSLs, and its impact on our ability to support the ``Engineering'' title in Software Languages Engineering.
This paper reports on a survey that quantitatively characterizes the description of experimental validation of DSLs in papers published in 15 of the most important scientific venues covering this research area, from 2001 to 2008. 

This paper is organized as follows. In section 2, we outline our review questions. In section 3, we present the research protocol followed in this systematic review of the current state of practice in SLE. In section 4, we discuss the inclusion and exclusion criteria in this review. In section 5, we present the main findings of our review. In section 6, we discuss those findings, the strengths, and weaknesses of the evidence collected for this review and their generalizability to the current state of practice. In section 7, we summarize the main conclusions and their implications to the SLE community.

\vspace*{-.25cm}
\section{Review questions}
\label{sec:ReviewQuestions}
\vspace*{-.35cm}
Our main motivation was to determine the extent to which the DSL community had presented evidence of its commitment to usability experimentation, in the context of proposals of new DSLs. In order to guide our systematic review on the state of practice, we start by stating our research questions:
\begin{itemize}
	\item Is there a concrete and detailed evaluation model to measure DSLs Usability?
	\item Is the DSL community concerned about experimental evaluation as a mechanism to prevent future problems emerging from the proposed DSLs?
	\item To what extent does the DSL community present evidence that the developed DSLs are easy to use and correspond to end-users needs?
\end{itemize}

In order to facilitate an objective and consistent view on each of the inspected papers, we broke these questions into more detailed criteria that we then used to classify the surveyed papers. These more detailed questions were:
\begin{itemize}
	\item \textbf{RQ1:} Does the paper report the development of a DSL?
	\item \textbf{RQ2:} Does the paper report the DSL development process with some detail?
	\item \textbf{RQ3:} Does the paper report any experimentation conducted for the assessment of the DSL?
	\item \textbf{RQ4:} Does the paper report the inclusion of end-users in the assessment of a DSL?
	\item \textbf{RQ5:} Does the paper report any sort of usability evaluation?
\end{itemize}

\vspace*{-.25cm}
\section{Review methods}
\label{sec:ReviewMethods}
\vspace*{-.25cm}
This paper reports on a survey that quantitatively characterizes the description of experimental validation of DSLs in papers published in 15 of the most important scientific publications covering this research area, from 2001 to 2008. 

The selected publications include: 1 special issue of a journal (\textit{Journal of Visual Languages and Computing (JVLC)}), 2 conferences (\textit{International Conference on Software Language Engineering (SLE)} and \textit{International Conference on Model Driven Engineering Languages and Systems (MODELS)}), and 10 workshop series focused in domain driven development and languages engineering, namely: \textit{IEEE Symposium on Visual Languages and Human-Centric Computing (VL/HCC)}, O\textit{OPSLA Workshop on Domain-Specific Modeling (DSM)}, \textit{OOPSLA Workshop on Domain-Specific Visual Languages (DSVL)}, \textit{ECOOP Workshop on Domain-Specific Program Development (DSPD)}, \textit{International Workshop on Language Engineering (ATEM)}, \textit{Model-Driven Development Tool Implementers Forum (MDD-TIF)}, \textit{Modellierung DSML (DSML)}, \textit{International Workshop on Software Factories at OOPSLA (OOPSLA-SF)}, \textit{ECOOP Workshop on Evolution and Reuse of Language Specifications for DSLs (ERLS)}, \textit{ETAPS Workshop on Language Descriptions}, \textit{Tools and Applications (LDTA)}. 
The survey also covers 2 general Software Engineering publications, namely \textit{IEEE Transactions on Software Engineering (TSE)} and the \textit{International Conference in Software Engineering (ICSE)} conference series. 

Paper selection was performed in two steps: (i) a direct inspection of paper abstracts and conclusions, to identify papers covering our research questions, followed by (ii) a full review of selected papers, to answer our research questions. This process was followed in 10 publications. The exceptions were Models, VL/HCC, LDTA, ICSE and TSE. In these venues, we started by using a web search mechanism to identify good candidates for further review, and then followed steps (i) and (ii) for those papers. The search keywords were ``domain-specific language'', ``domain-specific modeling'', ``DSL'', and ``DSM''.

Table 1
presents an overview of the selected papers. We grouped the publications in two categories, corresponding to the two paper selection strategies identified earlier.
Each table row presents the publication name, the number of available papers in that publication, from 2001 to 2008, the number of inspected papers, the number of selected papers and their percentage with respect to the number of inspected papers. 

\begin{table}
\label{tab:selectedpapers}
\vspace*{-.5cm}
\caption{Selected papers}
\vspace*{-.25cm}
\centering
\begin{tabular}{|l|l|l|l|l|l|}
\hline
\multirow{2}{*}{\textbf{Selection}} & \multirow{2}{*}{\textbf{Publication}} & \textbf{Available} & \textbf{Inspected} & \textbf{Selected} & \textbf{Selection} \\
 & & \textbf{articles} & \textbf{articles} & \textbf{articles} & \textbf{Percentage} \\
\hline
\multirow{9}{*}{\textbf{Direct}} & OOPSLA-DSM & 97 & 97 & 14 & 14.4\% \\ 
\cline{2-6}
 & OOPSLA-DSVL & 27 & 27 & 5 & 18.5\% \\ 
\cline{2-6}
 & DSPD & 19 & 19 & 3 & 15.8\% \\ 
\cline{2-6}
 & SLE & 18 & 18 & 0 & 0.0\% \\ 
\cline{2-6}
 & ATEM & 13 & 13 & 2 & 15.4\% \\ 
\cline{2-6}
 & MDD-TIF & 10 & 10 & 3 & 30.0\% \\ 
\cline{2-6}
 & DSML & 12 & 10 & 0 & 0.0\% \\ 
\cline{2-6}
 & OOPSLA-SF & 9 & 9 & 0 & 0.0\% \\ 
\cline{2-6}
 & ECOOP-ERLS & 6 & 6 & 0 & 0.0\% \\ 
\cline{2-6}
 & JVLC & 5 & 5 & 2 & 40.0\% \\ 
\hline
\multirow{5}{*}{\textbf{Query}} & VL/HCC & 141 & 16 & 2 & 12.5\% \\ 
\cline{2-6}
 & LDTA & 10 & 2 & 1 & 50.0\% \\ 
\cline{2-6}
 & MODELS & 200 & 4 & 1 & 25.0\% \\ 
\cline{2-6}
 & ICSE & 42 & 6 & 2 & 33.3\% \\ 
\cline{2-6}
 & TSE & 32 & 2 & 1 & 50.0\% \\ 
\hline
 & Total & 641 & 242 & 36 & 14.6\% \\ 
\hline
\end{tabular}
\vspace*{-.5cm}
\end{table}

Although we have used a common time frame for all publications, several of these publications were only available in some of the years under scrutiny.
This diversity of number of editions of the publications explains the variability of the number of scrutinized papers, with respect to their origin.
Nevertheless, we believe that our sample is representative of the current state of practice in DSL development.

The large majority of selected papers were published in workshops. This distribution can be regarded as an indicator of the relative novelty of this research area. 
Furthermore, our survey targeted publications with a strong concentration on discussions on DSLs, and their creation. However, the development of DSLs crosscuts the whole software industry. Therefore, it is fair to assume that many DSLs that are developed in this community are disseminated in journals, conferences, and workshops targeted to the domains they address.

\vspace*{-.25cm}
\section{Included and excluded studies}
\label{sec:IncludedExcluded}
\vspace*{-.25cm}
As discussed in section 3, the selection of eligible papers followed a two-step protocol. We started by a pre-selection, which was then followed by an in-depth analysis of each of the reviewed papers. We were conservative in the first step: when in doubt, papers went on to full review.

During the second step, we selected papers which would help answering each of our research questions. To facilitate paper selection, we defined strict paper inclusion criteria, namely:
(i) the paper reported on the development of at least one DSL;
(ii) the paper reported on the experimental evaluation of DSLs; or
(iii) the paper reported on specific techniques of DSLs Usability evaluation.
	
Criterion (i) helped establishing a baseline for developed DSLs which could be, or not, assessed through some sort of experimental evaluation. Criterion (ii) covered such evaluation. We also used criterion (iii) in our search, but had no success while applying it. This criterion was targeted at finding out trends with respect to specific techniques for usability testing that suit well the DSL community.

We excluded from our survey any papers not covering any of these issues. For instance, SLE includes, among other subjects, research on development frameworks for DSLs, generative programming, and model transformations to cope with the evolution of DSLs and its impact on the evolution of software built upon an evolving DSL. While these subjects are essential to the SLE research area, they are out of the scope of our survey.

A total of 36 papers were finally selected \cite{Barbero2007DSM,Bencomo2006DSM,Bencomo2008ICSE,Bettin2002DSVL,Bierhoff2006DSM,Carlson2001DSVL,Correal2007DSM,Evermann2005TSE,Furtado2006DSM,Grant2003DSM,Gray2001DSVL,Grundy2004JVLC,Haugen2007DSM,Hemel2008Models,Hosking2008VLHCC,Howard2002DSVL,Jouault2006DSPD,Kolovos2007MDDTIF,Luoma2004DSM,Merilinna2008DSM,Merilinna2007DSM,Mora2008DSM,Patki2008DSM,Pohjonen2007MDTIF,Prahofer2006DSM,Prahofer2007VLHCC,Reichert2008ICSE,Sadielek2007ATEM,Schmidt2002DSVL,Souza2008DSPD,Sprinkle2004JVLC,Svansson2007DSM,Tairas2007SLE,Teiken2008DSM,Trask2006DSPD,Zeng2006LDTA}.

\vspace*{-.25cm}
\section{Results}
\vspace*{-.25cm}
In this section, we report the obtained results, for each of our research questions.

\vspace*{-.25cm}
\subsubsection{RQ1: Does the paper report the development of a DSL?} 
It is important to understand which of the selected papers do report on the development of a DSL, so that we can later compute the percentage of such reports which include information on DSL evaluation.

Table 2
summarizes the relative weight of papers reporting DSL development among the selected papers. A considerable percentage of the total selected papers (91.7\%) reports the development of a DSL. Some of these DSLs were developed to satisfy a specific demand in the real world, while others were presented as a proof of concept, targeted to improve a specific domain in software production (e.g. a DSL for Interactive Television applications \cite{Barbero2007DSM,Kolovos2007MDDTIF,Pohjonen2007MDTIF}, or a DSL for interoperability between object-oriented and mainframe systems \cite{Souza2008DSPD}). In contrast, the 3 papers which do not report DSL development were selected for this survey because they covered topics which were relevant to it, namely the usage of quantitative analysis to assess domain-specific modelling techniques \cite{Bettin2002DSVL}, a survey on experiences while developing DSMLs \cite{Luoma2004DSM}, and  
a paper which refers to the usage of usability techniques for the assessment of DSLs \cite{Haugen2007DSM}.

\begin{table}
\label{tab:DSLDevelopment}
\vspace*{-.5cm}
\caption{Papers reporting DSL development}
\vspace*{-.25cm}
\centering
\begin{tabular}{|l|r|r|}
\hline
\textbf{DSL development} & \textbf{N} & \textbf{Percentage} \\ 
\hline
\textbf{Paper reports DSL development} & 33 & 91.7\% \\ 
\hline
\textbf{Paper does not report DSL development} & 3 & 8.3\%  \\ 
\hline
\multicolumn{1}{l}{} & \multicolumn{1}{l}{} & \multicolumn{1}{l}{} \\ 
\end{tabular}
\vspace*{-.5cm}
\end{table}

\vspace*{-.25cm}
\subsubsection{RQ2: Does the paper report the DSL development process with some detail?} This question aims to help us characterizing the extent to which authors provide details about the DSLs whose development was detailed in the papers. This information is relevant to our context, as an extra element for comparison. So, for each paper, we looked for details on the DSL construction. Out of the 33 papers reporting a DSL development, 16 provide some in-depth details on how those DSLs are built. The presence of a metamodel was not imperative, for this classification, but in some cases it proved to be a good help explaining the developed DSL. The distribution of papers reporting DSL construction details over the years is presented in table 3.

\begin{table}
\label{tab:DSLDevelopmentDetails}
\vspace*{-.5cm}
\caption{Papers reporting DSL development details}
\vspace*{-.25cm}
\centering
\begin{tabular}{|l|l|l|l|}
\hline
\textbf{Year} & \textbf{DSLs} & \textbf{Detailed Descriptions} & \textbf{Percentage} \\ 
\hline
\multicolumn{1}{|r|}{2008} & \multicolumn{1}{r|}{9} & \multicolumn{1}{r|}{4} & \multicolumn{1}{r|}{44\%} \\ 
\hline
\multicolumn{1}{|r|}{2007} & \multicolumn{1}{r|}{10} & \multicolumn{1}{r|}{7} & \multicolumn{1}{r|}{70\%} \\ 
\hline
\multicolumn{1}{|r|}{2006} & \multicolumn{1}{r|}{7} & \multicolumn{1}{r|}{2} & \multicolumn{1}{r|}{29\%} \\ 
\hline
\multicolumn{1}{|r|}{2005} & \multicolumn{1}{r|}{1} & \multicolumn{1}{r|}{1} & \multicolumn{1}{r|}{100\%} \\ 
\hline
\multicolumn{1}{|r|}{2004} & \multicolumn{1}{r|}{3} & \multicolumn{1}{r|}{2} & \multicolumn{1}{r|}{67\%} \\ 
\hline
\multicolumn{1}{|r|}{2003} & \multicolumn{1}{r|}{1} & \multicolumn{1}{r|}{0} & \multicolumn{1}{r|}{0\%} \\ 
\hline
\multicolumn{1}{|r|}{2002} & \multicolumn{1}{r|}{3} & \multicolumn{1}{r|}{0} & \multicolumn{1}{r|}{0\%} \\ 
\hline
\multicolumn{1}{|r|}{2001} & \multicolumn{1}{r|}{2} & \multicolumn{1}{r|}{0} & \multicolumn{1}{r|}{0\%} \\ 
\hline
\multicolumn{1}{|r|}{Totals} & \multicolumn{1}{r|}{36} & \multicolumn{1}{r|}{16} & \multicolumn{1}{r|}{44\%} \\ 
\hline
\end{tabular}
\vspace*{-.5cm}
\end{table}

Among the 3 papers which do not report the development of a DSL, two of them \cite{Luoma2004DSM,Haugen2007DSM} also discuss in some detail how DSLs are built. Overall, 16 out of the 36 selected papers provide this sort of details, and we can observe that the vast majority of the papers providing these details were published in the most recent half of the time frame considered in this survey.

\vspace*{-.25cm}
\subsubsection{RQ3: Does the paper report any experimentation conducted for the assessment of the DSL?} As we have seen so far, 33 of our selected papers report the development of a DSL. The next step is to find out how many of these papers report having performed any sort of experimental evaluation of the developed DSLs, and to characterize the reported experimentation. To achieve this, we will use two orthogonal categorizations of experimental validation. 

The first one is used to identify if the experimental validation reported by authors is of a \textbf{quantitative}, or a  \textbf{qualitative} nature. 
A Quantitative Method is based on the evaluation of measurable property (or properties) from real data, with the aim of supporting or refuting an hypothesis raised by the experimenter. Qualitative Methods focus on qualitative data obtained through observation, interviews, questionnaires, and so on, from a specific population. The data is then cataloged in such way that it can be useful to infer to other situations. In contrast with Quantitative Methods, no kind of measurable evaluation is performed. In spite of this apparent fragility, in many cases, qualitative methods may help to explain the reasons for some relationships and results, which otherwise would not be well understood \cite{Seaman1999TSE}.
As some of the papers claim that some sort of experimental validation is performed, but do not provide enough information for the reader to know which kind of evaluation was performed, we add a third category, labeled as \textbf{unknown}. Finally, some papers report no experimental validation at all. Table 4
summarizes our findings, using this classification.

\begin{table}
\label{tab:QuantitativeQualitative}
\vspace*{-.5cm}
\caption{Quantitative vs. Qualitative experimentation}
\vspace*{-.25cm}
\centering
\begin{tabular}{|l|l|l|}
\hline
\textbf{Experimentation kind} & \textbf{N} & \textbf{Percentage} \\ 
\hline
\textbf{Quantitative} & \multicolumn{1}{r|}{3} & \multicolumn{1}{r|}{8.3\%} \\ 
\hline
\textbf{Qualitative} & \multicolumn{1}{r|}{2} & \multicolumn{1}{r|}{5.6\%} \\ 
\hline
\textbf{Unknown} & \multicolumn{1}{r|}{21} & \multicolumn{1}{r|}{58.3\%} \\ 
\hline
\textbf{Without Experimentation} & \multicolumn{1}{r|}{10} & \multicolumn{1}{r|}{27.8\%} \\ 
\hline
\end{tabular}
\vspace*{-.5cm}
\end{table}

The first noticeable information is that only five of the papers are explicit about using a particular kind of experimental validation of DSLs. Two of the papers reporting on using quantitative experimentation do so within the scope of the description of a particular DSL. Zeng et al. report on comparison between the number of Lines of Code (LOC) generated by a DSL and the LOC when using a ``traditional'' approach to software development \cite{Zeng2006LDTA}. Merilinna et al. \cite{Merilinna2007DSM} also use a LOC-based comparison, among three development alternatives: ``traditional software development'', UML-based software development, and DSL-based software development. They use an atomic model element to measure the software production effort.
The third paper addressing experimental evaluation makes a comparison between two different approaches: the traditional software implementation versus using a DSML, in an effort to show DSMLs' benefits \cite{Bettin2002DSVL}. None of these papers presents data in such a manner that it could be reused in other experiments.

One of the papers \cite{Correal2007DSM} reports the usage of a qualitative approach to the assessment of DSM techniques targeted to the definition and improvement of software process models within a software development company. The data presented in this study is also not well suited to facilitate a meta-analysis. The other paper covering qualitative assessment \cite{Luoma2004DSM}, reports a 20 industrial project research using DSMs and MetaEdit+. The qualitative data was collected through diverse means: interviews and discussions with consultants or in-house developers who created the DSMLs, with domain engineers, responsible personnel for the architectural solution and tool support.

In contrast, 10 papers do not report the implementation of any kind of experimental evaluation of DSLs \cite{Bencomo2006DSM,Bencomo2008ICSE,Grant2003DSM,Grundy2004JVLC,Haugen2007DSM,Hosking2008VLHCC,Prahofer2006DSM,Prahofer2007VLHCC,Tairas2007SLE,Teiken2008DSM}.

21 papers share two common features: (i) they do not provide information concerning whether the work they describe involved any sort of experimental validation, but (ii) they do provide some information concerning the kinds of examples used in such validation. 20 of these papers use ad-hoc, or toy examples \cite{Barbero2007DSM,Bierhoff2006DSM,Carlson2001DSVL,Evermann2005TSE,Furtado2006DSM,Gray2001DSVL,Hemel2008Models,Howard2002DSVL,Jouault2006DSPD,Kolovos2007MDDTIF,Merilinna2008DSM,Mora2008DSM,Patki2008DSM,Pohjonen2007MDTIF,Reichert2008ICSE,Sadielek2007ATEM,Schmidt2002DSVL,Souza2008DSPD,Sprinkle2004JVLC,Svansson2007DSM,Trask2006DSPD}, while 1 claims to have obtained their information at an industrial level \cite{Luoma2004DSM}, but does not provide details on the particular evaluation. 2 papers \cite{Correal2007DSM,Luoma2004DSM} use industry-level examples, while the remaining 3 \cite{Bettin2002DSVL,Merilinna2007DSM,Zeng2006LDTA} provide no details concerning the kind of examples used in their validation. Table 5 summarizes this information.

\begin{table}
\label{tab:ToyIndustry}
\vspace*{-.25cm}
\caption{Toy vs. industrial examples usage}
\vspace*{-.25cm}
\centering
\begin{tabular}{|l|l|l|}
\hline
\textbf{Experimental material kind} & \textbf{N} & \textbf{Percentage} \\ 
\hline
\textbf{Ad-hoc/Toy example} & \multicolumn{1}{r|}{21} & \multicolumn{1}{r|}{58.3\%} \\ 
\hline
\textbf{Industrial level} & \multicolumn{1}{r|}{2} & \multicolumn{1}{r|}{5.6\%} \\ 
\hline
\textbf{Unknown} & \multicolumn{1}{r|}{3} & \multicolumn{1}{r|}{8.3\%} \\ 
\hline
\textbf{Without Experimentation} & \multicolumn{1}{r|}{10} & \multicolumn{1}{r|}{27.8\%} \\ 
\hline
\end{tabular}
\vspace*{-.5cm}
\end{table}

\vspace*{-.75cm}
\subsubsection{RQ4: Does the paper report the inclusion of end-users in the assessment of a DSL?}
When developing a new system, regardless of it being completely new or a new system version, it is important to know its' intended users profile, and how they will use the system. Their impact on usability is enormous, so an early definition of their capabilities allows developers to understand what is important and what is disposable, reducing the number of redundant or unnecessary features in the system \cite{Nielson1993}. This observation is applicable to software users in general, and to DSL users in particular. In order to characterize the DSL users who participate in a DSL evaluation, we define three categories:
(i)\textbf{Industrial or specialized personnel} - we use this classification for papers reporting subjects with expertise in the domain. The domain expert does not necessarily need to have knowledge about DSLs, in general;
(ii) \textbf{Academic} - the typical example is the usage of graduate students as surrogates for the real end-users of a DSL;
(iii) \textbf{Not defined} - we use this category whenever we are not able to find the user profile in the paper.
  
3 papers reported using domain experts, including seismologists \cite{Sadielek2007ATEM} and other specialized developers \cite{Correal2007DSM,Luoma2004DSM}. The remaining 2 papers \cite{Hosking2008VLHCC,Zeng2006LDTA} did not specify the type of subjects involved in the evaluation of the DSL. 

\begin{table}
\label{tab:DomainExperts}
\vspace*{-.5cm}
\caption{Domain experts usage}
\vspace*{-.25cm}
\centering
\begin{tabular}{|l|l|l|}
\hline
\textbf{Domain experts usage} & \textbf{N} & \textbf{Percentage} \\ 
\hline
\textbf{Industrial or specialized personnel} & \multicolumn{1}{r|}{3} & \multicolumn{1}{r|}{8.3\%} \\ 
\hline
\textbf{Academic} & \multicolumn{1}{r|}{0} & \multicolumn{1}{r|}{0.0\%} \\ 
\hline
\textbf{Not defined} & \multicolumn{1}{r|}{2} & \multicolumn{1}{r|}{5.6\%} \\ 
\hline
\textbf{Unknown} & \multicolumn{1}{r|}{31} & \multicolumn{1}{r|}{86.1\%} \\ 
\hline
\end{tabular}
\vspace*{-.5cm}
\end{table}

\vspace*{-.75cm}
\subsubsection{RQ5: Does the paper report any sort of usability evaluation?}
Usability is a quality attribute based on users' and/or stakeholders' needs satisfaction by assessing how easy a system is to use. We assess the extent to which DSLs were tested for usability and whether they fulfill the end user needs, and identify three categories for this:
(i) \textbf{Usability Techniques} - the papers report a set of techniques that allow DSLs becoming more accurate to the end users;
(ii) \textbf{Ad-hoc} - the paper reports an ad-hoc approach to improving DSLs' usage without a detailed rationale;
(iii) \textbf{No usability evaluation} - the paper provides no usability evaluation.
Table 7
summarizes this information.

\begin{table}
\label{tab:Usability}
\vspace*{-.5cm}
\caption{Reported Usability techniques in DSL validation}
\vspace*{-.25cm}
\centering
\begin{tabular}{|l|r|r|}
\hline
\textbf{Usability Approach} & \textbf{N }& \textbf{Percentage} \\ 
\hline
\textbf{Usability techniques} & 1 & 2.8\% \\ 
\hline
\textbf{Ad-hoc} & 6 & 16.6\% \\ 
\hline
\textbf{No usability evaluation} & 29 & 80.6\% \\ 
\hline
\end{tabular}
\vspace*{-.5cm}
\end{table}

The paper that used Usability Techniques has adapted techniques for general purpose languages to the DSLs' context \cite{Haugen2007DSM}. Papers in the Ad-hoc category focused on visual issues pointed out by subjects, such as layout \cite{Hosking2008VLHCC}, usage of familiar icons and commands \cite{Mora2008DSM}, interactive dialogs to increase users performance \cite{Schmidt2002DSVL}, and the impact that an iterative development process in cooperation with subjects has in usability \cite{Sadielek2007ATEM}. Finally, \cite{Grundy2004JVLC} developed three Domain Specific Visual Languages, each one with an intended target user group, and reported using usability trials, without specifying the exact procedure. The author also compared several parameters, such as consistency and error-proneness. 

\vspace*{-.25cm}
\section{Discussion}
\vspace*{-.25cm}
We found a low level of experimentation reported in the surveyed papers. Although roughly half of the papers report with some detail the development process of the DSL, only about 14\% of the papers report either a quantitative or a qualitative evaluation of the DSL and they provide very few details on what was done. Researchers planning to replicate such evaluations would suffer from a lot of tacit knowledge, which is a well-known factor hampering the independent validation of claims supported through experimentation\cite{Shull2002ISESE,Shull2004ESE}. The proposal of a roadmap for the validation of DSLs could mitigate this shortcoming of current practice. A widely accepted methodology for DSLs validation would be helpful for researchers and practitioners.

A shortcoming in the reviewed work was the predominance of toy examples, when compared to the usage of industry level examples. This represents a threat to the validity of claims made in such papers, as the conclusions drawn from toy examples do not necessarily scale up to industry. 
Most of the publications scrutinized in this review are workshops. Therefore, it may be the case that 
the predominance of work in progress papers in such venues 
increases the relative frequence of insuficiently validated claims. 

The lack of detail on the surveyed experimentation reports implies that we often do not know who were the subjects involved in the process. This is a threat, as we do not know the extent to which domain experts were really involved in this process, in most cases. Without characterizing the users, the validity of any conclusions concerning the usability of the DSLs is completely questionable. For instance, a newbie might have difficulties using a DSL, not because of problems with the DSL itself, but due to shortcomings of his own expertise in the domain the DSL is targeted to.

In any systematic review, we must explicitly address its inherent validity threats \cite{Kitchenham2007SR}. 
Although we were very conservative in our selection (when in doubt, we kept the paper for further review), it is always possible that some papers may have been missed, either because we failed to understand the abstract, or because the abstract was incomplete and did not cover the validation of the proposals with enough detail. Another common threat in systematic reviews concerns the misclassification of papers. This can happen when the reviewers mis-understand some important information about the paper and classify it in the wrong category (e.g., a qualitative study is counted as a quantitative one). We mitigated this threat by creating objective criteria to classify the surveyed papers, thus minimizing subjectiveness in the data collection.

\vspace*{-.25cm}
\section{Conclusions}
\vspace*{-.3cm}
Domain driven development is an increasingly popular way of developing software that aims to leverage the contribution of domain experts in such development. There are several claims of the benefits of this approach, in well-defined and constrained domains. However, we did not find evidence supporting such claims, either because it is not made explicit, or because it does not exist.

The SLE community does not systematically report on the realization of any sort of experimental validation of the languages it builds. While this does not necessarily mean that no evaluation is performed, it sends the wrong message to Engineering practitioners, which should always be concerned in systematically evaluating its products. In this paper, we support this claim by reviewing a large set of publications from the SLE community. 

One of the outcomes from this work is that now the SLE community has an evidence to back up the awareness to this problem, which is necessary first step toward solving it. Therefore, one of the present challenges to the  community, which is clearly interested in reinforcing the ``Engineering'' in SLE, is to foster the systematic evaluation of the produced languages as part of the standard of practice in the development process. The solution may either to go for a \textit{``de facto''} solution, based on a high standard state of practice (this would require the community to be concerned with this subject, when publishing its work), or for a ``the jure'' solution, in which the community would set up an agreed standard for enforcing this kind of validation. 

\vspace*{-.25cm}
\section*{Acknowledgements}
\vspace*{-.3cm}
The authors would like to thank Juha-Pekka Tolvanen for his valuable comments on a draft of this paper and CITI for the support for this research.
\vspace*{-.3cm}
\bibliography{Bibliography}
\bibliographystyle{splncs}

\end{document}